\documentclass{IEEEtran}
\usepackage{booktabs}
\usepackage{cite}
\usepackage{amsmath,amssymb,amsfonts,amstext}% Lots of math symbols and enviro
\usepackage{graphicx} % For including graphics N.B. pdftex graphics dr
\usepackage{algorithm}
\usepackage{algorithmic}
\usepackage{multirow}
\usepackage{caption}
\usepackage{stfloats}
\usepackage{url}
\usepackage{bbm}

\DeclareMathOperator*{\mini}{minimize}
\DeclareMathOperator*{\maxi}{maximize}
\DeclareMathOperator{\sbto}{subject \text{ } to}

\newtheorem{theorem}{Theorem}[section]

\newtheorem{proposition}[theorem]{Proposition}

\newcommand{\qed}{\nobreak \ifvmode \relax \else
      \ifdim\lastskip<1.5em \hskip-\lastskip
      \hskip1.5em plus0em minus0.5em \fi \nobreak
      \vrule height0.75em width0.5em depth0.25em\fi}

\newcounter{MYtempeqncnt}

\ifCLASSINFOpdf
\else
\fi
\usepackage{fixltx2e}
\usepackage{stfloats}

\begin{document}
%
% paper title
% can use linebreaks \\ within to get better formatting as desired
\title{Sparse Beamforming and User-Centric Clustering for Downlink Cloud Radio Access Network}

% author names and affiliations
% use a multiple column layout for up to three different
% affiliations
\author{\IEEEauthorblockN{Binbin Dai, \IEEEmembership{Student Member, IEEE} and Wei Yu, \IEEEmembership{Fellow, IEEE}}
%\IEEEauthorblockA{Department of Electrical and Computer Engineering\\
%         University of Toronto, Toronto, Ontario M5S 3G4, Canada  \\
%				Emails: \{bdai, weiyu\}@comm.utoronto.ca}
%\and
%\IEEEauthorblockN{Homer Simpson}
%\IEEEauthorblockA{Twentieth Century Fox\\
%Springfield, USA\\
%Email: homer@thesimpsons.com}
%\and
%\IEEEauthorblockN{James Kirk\\ and Montgomery Scott}
%\IEEEauthorblockA{Starfleet Academy\\
%San Francisco, California 96678-2391\\
%Telephone: (800) 555--1212\\
%Fax: (888) 555--1212}
\thanks{Manuscript accepted and to appear in IEEE Access, Special Issue on Recent Advances in Cloud Radio Access Networks, 2014. 
The materials in this paper have been presented in part at 
the IEEE International Workshop on Signal Processing Advances in 
Wireless Communications (SPAWC), Toronto, Canada, June 2014, \cite{binbin14}.
This work was supported by Huawei Technologies Canada and by
Natural Sciences and Engineering Research Council (NSERC) of Canada.
The authors are with The Edward S. Rogers Sr. Department of Electrical
and Computer Engineering, University of Toronto, Toronto, ON M5S 3G4, 
Canada. (e-mails: bdai@ece.utoronto.ca, weiyu@comm.utoronto.ca).}
}
% conference papers do not typically use \thanks and this command
% is locked out in conference mode. If really needed, such as for
% the acknowledgment of grants, issue a \IEEEoverridecommandlockouts
% after \documentclass

% for over three affiliations, or if they all won't fit within the width
% of the page, use this alternative format:
% 
%\author{\IEEEauthorblockN{Michael Shell\IEEEauthorrefmark{1},
%Homer Simpson\IEEEauthorrefmark{2},
%James Kirk\IEEEauthorrefmark{3}, 
%Montgomery Scott\IEEEauthorrefmark{3} and
%Eldon Tyrell\IEEEauthorrefmark{4}}
%\IEEEauthorblockA{\IEEEauthorrefmark{1}School of Electrical and Computer Engineering\\
%Georgia Institute of Technology,
%Atlanta, Georgia 30332--0250\\ Email: see http://www.michaelshell.org/contact.html}
%\IEEEauthorblockA{\IEEEauthorrefmark{2}Twentieth Century Fox, Springfield, USA\\
%Email: homer@thesimpsons.com}
%\IEEEauthorblockA{\IEEEauthorrefmark{3}Starfleet Academy, San Francisco, California 96678-2391\\
%Telephone: (800) 555--1212, Fax: (888) 555--1212}
%\IEEEauthorblockA{\IEEEauthorrefmark{4}Tyrell Inc., 123 Replicant Street, Los Angeles, California 90210--4321}}

% use for special paper notices
%\IEEEspecialpapernotice{(Invited Paper)}

% make the title area
\maketitle

\begin{abstract}

This paper considers a downlink cloud radio access network (C-RAN) in which all the base-stations (BSs) are connected 
to a central computing cloud via digital backhaul links with finite capacities. 
Each user is associated with a user-centric cluster of BSs; the central processor shares the user's data with the BSs in the cluster, 
which then cooperatively serve the user through joint beamforming.
Under this setup, this paper investigates the user scheduling, BS clustering and beamforming design problem from a network utility maximization perspective.
Differing from previous works, this paper explicitly considers the per-BS backhaul capacity constraints. 
We formulate the network utility maximization problem 
for the downlink C-RAN under two different models depending on whether the BS clustering 
for each user is \emph{dynamic} or \emph{static} over different user scheduling time slots.
In the former case, the user-centric BS cluster is dynamically optimized for each scheduled user along 
with the beamforming vector in each time-frequency slot,  
while in the latter case the user-centric BS cluster is fixed for each user and 
we jointly optimize the user scheduling and the beamforming vector to account for the backhaul constraints. 
In both cases, the nonconvex per-BS backhaul constraints are approximated using the reweighted $\ell_1$-norm technique.
This approximation allows us to reformulate the per-BS backhaul constraints into \emph{weighted} per-BS power constraints 
and solve the weighted sum rate maximization problem through a generalized weighted minimum mean square error approach.
This paper shows that the proposed dynamic clustering algorithm can achieve
significant performance gain over existing naive clustering schemes. 
This paper also proposes two heuristic static clustering schemes that 
can already achieve a substantial portion of the gain.

\end{abstract}
% IEEEtran.cls defaults to using nonbold math in the Abstract.
% This preserves the distinction between vectors and scalars. However,
% if the conference you are submitting to favors bold math in the abstract,
% then you can use LaTeX's standard command \boldmath at the very start
% of the abstract to achieve this. Many IEEE journals/conferences frown on
% math in the abstract anyway.


% For peer review papers, you can put extra information on the cover
% page as needed:
% \ifCLASSOPTIONpeerreview
% \begin{center} \bfseries EDICS Category: 3-BBND \end{center}
% \fi
%
% For peerreview papers, this IEEEtran command inserts a page break and
% creates the second title. It will be ignored for other modes.
\IEEEpeerreviewmaketitle

\section{Introduction}

%\IEEEPARstart{C}{urrent} cellular network has been deployed with ultra-dense small cells in order to 
%support the exponentially increasing demand of data traffic. 

\IEEEPARstart{T}{he} fifth-generation (5G) wireless system is expected to support an ever increasing number of 
mobile devices with ubiquitous service access. 
To realize this 5G vision, ultra-dense small cell deployments and cloud computing 
are recognized as the two key enabling technologies \cite{Rost14}.
With small cells, the received signal strength is enhanced at the user's side due to 
the reduced distance to the serving base-stations (BSs). 
However, as the neighboring BSs are also located closer in distance, the users are exposed to 
more inter-cell interference, which limits the performance of the cellular network.
Cloud radio access network (C-RAN) is an emerging network architecture that is capable of dealing with this 
inter-cell interference issue. 
In C-RAN, the BSs are connected to a central processor (CP) via digital backhaul links.
This allows the CP to jointly encode the user messages using linear precoding or beamforming techniques 
for interference mitigation purpose in the downlink.
The C-RAN architecture can be thought of as a platform for the 
practical implementation of network multiple-input multiple-output (MIMO) and
coordinated multi-point (CoMP) transmission concepts \cite{Wei10}.

This paper studies the optimization of the C-RAN architecture focusing on the effect of finite-capacity backhaul links 
on the overall network capacity.
In this realm, several practical transmission strategies have been proposed for the downlink C-RAN to account for the finite backhaul.
In one such strategy, the CP performs the beamforming operation, then compresses and forwards the beamformed signals 
to the BSs. 
Compression is needed because of the capacity limits of the backhaul links. 
This strategy is investigated in \cite{simeone2009, Park13} and is referred to as the \emph{compression} strategy in this paper.
In an alternative strategy, each user is associated with a cluster of multiple BSs 
and the CP simply shares each user's message directly with its serving BS cluster.
The BSs form the beamformed signals locally, then cooperatively transmit the signals to the users. 
This strategy is studied in \cite{simeone2009, marsch2008, Gesbert11} 
and is referred to as the \emph{data sharing} strategy in this paper. 
In the compression strategy, the amount of available backhaul capacity determines the resolution of the compressed signals: 
higher-resolution compression requires larger backhaul capacity.
In the data sharing strategy, the amount of required backhaul capacity is related to the BS cluster size:
larger cluster size leads to higher backhaul consumption.

This paper focuses on the data sharing strategy. % since it is relatively easy to implement in practice.
The performance of the data sharing strategy depends crucially on the choice of BS cooperation cluster for each user. 
Broadly speaking, there are two types of 
BS clustering schemes for data sharing: \emph{disjoint clustering} and \emph{user-centric clustering}.
In disjoint clustering scheme, the entire network is divided into non-overlapping clusters and the BSs in each cluster
jointly serve all the users within the coverage area \cite{CoMP11}. 
Although disjoint clustering scheme has already been shown to be effective 
in mitigating the inter-cell interference \cite{Huang07, Huang09}, 
users at the cluster edge still suffer from considerable inter-cluster interference.
Differently, in user-centric clustering, each user is served by an individually selected subset of neighboring BSs and 
different clusters for different users may overlap. 
The benefit of user-centric clustering is that there exists no explicit cluster edge. 
This paper adopts the user-centric clustering scheme and further considers two different implementations of user-centric clustering depending on whether BS clustering is \emph{dynamic} or \emph{static} over the different user scheduling time slots.
In dynamic clustering, the BS cluster for each user can change over time, allowing for more freedom to fully utilize the 
backhaul resources. 
However, dynamic clustering scheme also requires more signaling overhead as new BS-user associations need to be established continuously. 
In static clustering, the BS-user association is fixed over time and may only need to be updated as the user location changes.

This paper considers both dynamic and fixed clustering schemes, and proposes joint clustering, user scheduling and beamforming designs for the downlink C-RAN with user-centric data sharing strategy. We explicitly take per-BS backhaul capacity constraints into account in the network utility maximization framework, and use the $\ell_1$-norm reweighting technique in compressive sensing and a generalized weighted minimum mean square error (WMMSE) \cite{Cioffi08, Tom11} approach to solve the problem. We show that dynamic clustering can significantly outperform the naive channel strength based clustering strategy, while the proposed heuristic static clustering schemes can already achieve a substantial portion of the performance gain.

\subsection{Related Work}

The information-theoretical capacity of the C-RAN model has been considered extensively in the literature. 
However, most of the theoretical analysis on C-RAN is restricted to 
simplified channel models \cite{Shamai08, Jiang12, marsch2008, Gesbert11}. 
Specifically, the achievable rate regions derived in \cite{marsch2008, Gesbert11} are based on a two-BS-two-user channel model, while 
\cite{Shamai08} and \cite{Jiang12} consider Wyner-like channel models and report the  
achievable rates and capacity bounds, respectively.
In \cite{Hoon12, Rong12}, large-system analysis of network MIMO system is carried out.
Although based on simplified models, 
these previous information-theoretical results already reveal the benefits of C-RAN in significantly improving the system performance.

This paper focuses on practical system design for the downlink C-RAN.
This design problem has been considered in the literature under various performance metrics. 
For instance, under the signal-to-interference-and-noise ratio (SINR) constraints at the receivers, 
\cite{Zhao12} considers backhaul minimization while \cite{Shi13, Rui14} consider network power minimization as the objective. 
Furthermore, 
the optimal tradeoff between the backhaul capacity and transmit power is investigated in \cite{binbin13, Zhuang14, Vu14}, 
while several other performance measures like mean square error (MSE) and energy efficiency (bits/Joule delivered to the users) are  considered in \cite{Shi08, Park12} and \cite{Sarkiss12, Ng13}, respectively.

In this paper, we consider the network utility as the performance measure for the downlink C-RAN.
Differing from the utility maximization problems in conventional wireless networks with only transmit power constraints, 
the additional backhaul constraints in C-RAN make the problem more challenging as the backhaul 
consumption at a particular BS is a function of not only the (continuous) user
rates but also the (discrete) number of associated users.
To tackle this mixed continuous and discrete optimization problem, existing literature mostly 
take the limited backhaul capacities into account \emph{implicitly} either by fixing the BS clusters \cite{Jun09, Ng10, Kaviani12}
or by adding the backhaul as a penalized term into the objective function \cite{hong12, Chowdhery11, Shervin2012}. 
Specifically, \cite{Jun09} considers sum rate maximization under fixed and disjoint clustering scheme while
\cite{Ng10} and \cite{Kaviani12} maximize a more general utility function under user-centric but predetermined BS clusters.
Dynamic user-centric clustering design is considered in \cite{hong12} by penalizing
the objective function with 
an $\ell_2$-norm approximation of the cluster size.
Alternatively, \cite{Chowdhery11} and \cite{Shervin2012} choose the backhaul rate as the penalized term but 
solve the problem heuristically.
For fixed clustering scheme assumed in \cite{Jun09, Ng10, Kaviani12}, the backhaul consumption is only known
afterwards by evaluating the rates of user messages delivered in each backhaul link.
For dynamic clustering designs considered in \cite{hong12, Chowdhery11, Shervin2012}, one has to optimally choose the 
price associated with each penalized term to ensure that the overall backhaul stays within the budget, which is not easy.

In contrast to all the above existing works in network utility maximization for the downlink C-RAN, this paper \emph{explicitly} 
formulates the per-BS backhaul constraints in the optimization framework.
With explicit per-BS backhaul constraints, we show that the backhaul resources can be 
more efficiently utilized and that the network utility can be significantly improved.

The BS clustering problem for C-RAN with limited backhaul capacity is combinatorial in nature, 
for which finding the global optimum is expected to be quite challenging. 
Several suboptimal cluster formation algorithms have already been proposed in the literature.
In \cite{Baracca12}, the BS cluster is assumed to be selected from a set of predetermined candidate clusters, and a greedy
cluster selection algorithm is proposed to maximize the network utility.
Alternatively, \cite{Moon11} models the cluster formation problem using graph theory, while \cite{Ying11} treats the problem from 
queuing theory perspective. 

This paper differs from previous work in that we propose a dynamic clustering scheme by optimizing a 
\emph{sparse} beamforming vector for each user, where the nonzero entries in the beamforming vector correspond to the user's
serving cluster. 
This allows the formulations of the BS clustering problems as an $\ell_0$-norm optimization problem and 
its subsequent solution via an application of the $\ell_1$-norm reweighting technique in compressive sensing.

This paper proposes a novel application of WMMSE approach to jointly optimize the user scheduling and beamforming vectors under
either dynamic or fixed BS clustering. 
This is in contrast to \cite{Ng10}, which uses the first-order Taylor expansion to approximate the 
the nonconvex rate expression.
Although the WMMSE approach has been applied to the C-RAN setup in the past \cite{Kaviani12, hong12},
these previous works do not explicitly take the backhaul constraints into consideration. 
As related work, the WMMSE approach has also been adapted to solve the max-min fairness problem for MIMO interfering 
broadcast channel \cite{Razaviyayn13}, a link flow rate control problem for the radio access network \cite{Liao14} and
a power minimization problem under time-averaged user rate constraints for the CoMP architecture \cite{Han14}. 
Recently, the WMMSE technique is generalized in \cite{Hong13} to a wider class of network setups 
using the successive convex approximation idea.
Finally, we mention that the WMMSE is a numerical approach for solving for a stationary point to the weighted sum rate (WSR) maximization 
problem over the beamformers, which is known to be a challenging problem. 
Recent progress for finding globally optimal solution to the WSR maximization problem under various conditions has been 
reported in \cite{TanFriedland11, Tan11, Lai14}.

\subsection{Main Contributions}

This paper considers the user scheduling, user-centric BS clustering and beamforming design problem for the downlink C-RAN.
The main contributions in this paper are summarized as follows:
\begin{enumerate}
\item Per-BS backhaul constraints are explicitly considered in the network utility maximization problem for 
the downlink C-RAN under data sharing strategy. 
A key novel technique proposed in this paper is that the per-BS backhaul constraint can be formulated in a weighted $\ell_0$-norm format and approximated using the reweighted $\ell_1$-norm.

\item A novel application of the WMMSE approach is proposed to solve the utility maximization problem with backhaul constraints. 
The proposed algorithm can be applied to the cases where the BS clustering for each user can be either dynamic or static.

\item We show numerically that with explicit per-BS backhaul constraints, the proposed algorithm is able to utilize the backhaul resources more efficiently, as well as to offer more flexibilities in choosing the cluster size.
Simulation results also show that as compared with the naive clustering schemes, both the dynamic and the static clustering 
schemes proposed in this paper achieve significant performance improvement.

\end{enumerate}

\subsection{Paper Organization and Notations}

The rest of the paper is organized as follows.
Section II introduces the system model.
Section III considers dynamic BS clustering and proposes a joint scheduling, beamforming
and clustering design algorithm together with two additional techniques to further reduce the computational complexity of the proposed 
algorithm.
In Section IV, the user scheduling and beamforming vectors are jointly optimized under fixed BS clustering and 
two heuristic static clustering algorithms are proposed. 
Numerical results are provided in Section V. 
Conclusions are drawn in Section VI.

Throughout this paper, lower-case bold letters (e.g. $\mathbf{w}$) denote vectors and upper-case bold letters (e.g. $\mathbf{H}$) denote matrices. We use $\mathbb{R}$ and $\mathbb{C}$ to denote real and complex domain, respectively. The matrix inverse, conjugate transpose and $\ell_p$-norm of a vector are denoted as $(\cdot)^{-1}$, $(\cdot)^H$ and $|\cdot|_p$ respectively. 
The complex Gaussian distribution is represented by $\mathcal{CN}(\cdot,\cdot)$ while $\text{Re}\{\cdot\}$ stands for the real part of a scalar. 
The expectation of a random variable is denoted as $\mathsf{E} \left [ \cdot \right]$. 
Calligraphy letters are used to denote sets while $|\cdot|$ stands for either the size of a set or the absolute value of a real scalar, depending on the context.

% no \IEEEPARstart
%This demo file is intended to serve as a ``starter file''
%for IEEE conference papers produced under \LaTeX\ using
%IEEEtran.cls version 1.7 and later.
%% You must have at least 2 lines in the paragraph with the drop letter
%% (should never be an issue)
%I wish you the best of success.
%
%\hfill mds
 %
%\hfill January 11, 2007
%
%\subsection{Subsection Heading Here}
%Subsection text here.
%
%
%\subsubsection{Subsubsection Heading Here}
%Subsubsection text here.

% An example of a floating figure using the graphicx package.
% Note that \label must occur AFTER (or within) \caption.
% For figures, \caption should occur after the \includegraphics.
% Note that IEEEtran v1.7 and later has special internal code that
% is designed to preserve the operation of \label within \caption
% even when the captionsoff option is in effect. However, because
% of issues like this, it may be the safest practice to put all your
% \label just after \caption rather than within \caption{}.
%
% Reminder: the "draftcls" or "draftclsnofoot", not "draft", class
% option should be used if it is desired that the figures are to be
% displayed while in draft mode.
%
%\begin{figure}[!t]
%\centering
%\includegraphics[width=2.5in]{myfigure}
% where an .eps filename suffix will be assumed under latex, 
% and a .pdf suffix will be assumed for pdflatex; or what has been declared
% via \DeclareGraphicsExtensions.
%\caption{Simulation Results}
%\label{fig_sim}
%\end{figure}

% Note that IEEE typically puts floats only at the top, even when this
% results in a large percentage of a column being occupied by floats.

\section{System Model}

\begin{figure}[t]
  \centering
  \includegraphics[width= 0.45\textwidth]{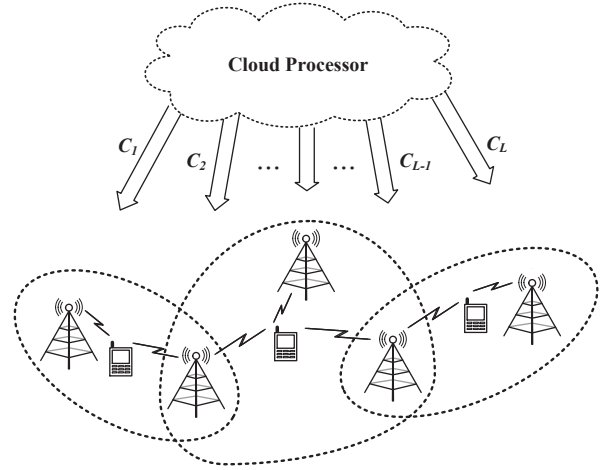}
\caption{Downlink C-RAN with per-BS backhaul capacity limits, where each user is cooperatively served by a user-centric and potentially
overlapping subset of BSs.}
\label{fig:SystemModel}
\end{figure}

Consider a downlink C-RAN with $L$ BSs and $K$ users, 
where each BS has $M$ transmit antennas while each user has $N$ receive antennas. 
Each BS $l$ is connected to a CP with a backhaul link with capacity limit $C_l, l \in \mathcal{L}=\left\{1,2,\cdots, L\right\}$,
as depicted in Fig.~\ref{fig:SystemModel}. 
We assume that the CP has access to all users' data
and distributes each user's data to an individually selected cluster of BSs via the backhaul links. 
Each user is then cooperatively served by its serving cluster
through joint beamforming.

%The transmit strategy design for each user in C-RAN mainly consists of three components:
%\emph{serving cluster selection}, \emph{transmit beamforming design} and \emph{transmit power allocation}.
In order to represent the BS cluster and transmit beamformer in a compact form, we 
introduce a network-wide beamforming vector
$\mathbf{w}_k = [\mathbf{w}_{k}^1,
\mathbf{w}_{k}^2, \cdots, \mathbf{w}_{k}^L] \in \mathbb{C}^{M_t \times 1} $ for each user $k \in \mathcal{K}=\left\{1,2,\cdots,K\right\}$, where $M_t = LM$ and $\mathbf{w}_{k}^l \in
\mathbb{C}^{M \times 1}$ is the transmit beamformer from BS $l$ to user $k$. 
Suppose that BS $l$ is not part of user $k$'s serving cluster, then the corresponding beamformer block
$\mathbf{w}_{k}^l$ is set to $\mathbf{0}$. 
Since each user is expected to be served by only a small number of BSs, the network-wide beamforming 
vector $\mathbf{w}_{k}$ is group sparse.
Here, we assume that all the $L$ BSs can potentially serve each scheduled user in order to simplify the notations.
However, the proposed algorithms in this paper
can be readily applied to the situation where only a subset of BSs are considered
as each user's candidate serving BSs\footnote{In the simulation part of this paper, only the strongest few BSs around each user
are considered as the candidate serving BSs in order to reduce the computational complexity of the proposed algorithms.}.

With linear transmit beamforming scheme at the BSs, the received
signal at user $k$, denoted as $\mathbf{y}_k \in \mathbb{C}^{N \times 1}$, can be written as
\begin{equation}
\mathbf{y}_k = \mathbf{H}_k \mathbf{w}_k s_k + \sum_{j \neq k, j \in \mathcal{K}} \mathbf{H}_k \mathbf{w}_j s_j  +\mathbf{n}_k, 
\label{eq:newYk}
\end{equation}
where $\mathbf{H}_k \in \mathbb{C}^{N \times M_t}$ denotes the channel state information (CSI) matrix from 
all the $M_t$ transmit antennas to user $k$, $\mathbf{n}_k \in \mathbb{C}^{N \times 1}$ is the received noise at user $k$ and is assumed to be distributed as $\mathcal{CN}(0, \sigma^2 \mathbf{I})$.
In this paper, we consider
the case where each user has only a single data stream for simplicity and assume that user $k$'s message $s_k$
is independent and identically distributed according to $\mathcal{CN}(0,1)$.
Under this consideration, the achievable rate for user $k$ can be written as:
\begin{align} \label{eq:Rk2}
&R_k = \\ \nonumber
& \log \left( 1 +\mathbf{w}_{k}^{H} \mathbf{H}_{k}^{H} \left (\sum_{\substack{j \neq k \\ j \in \mathcal{K}}} \mathbf{H}_{k} \mathbf{w}_j \mathbf{w}_{j}^{H} \mathbf{H}_{k}^{H} + \sigma^2 \mathbf{I} \right )^{-1} \mathbf{H}_{k} \mathbf{w}_k \right) \nonumber .
\end{align}

Note that the rate expression (\ref{eq:Rk2}) can also account for the user scheduling operation. 
A user $k$ is scheduled, i.e. $R_k$ is nonzero, if and only if its beamformer vector $\mathbf{w}_{k}$ is nonzero. In other words, the 
scheduling choice is determined by the indicator function:
\begin{equation}\label{eq:indicator}
 \mathbbm{1}\left\{ \left\|\mathbf{w}_{k}\right\|_2^{2} \right\}  = \left \{
   \begin{array}{l}
	 0, \quad \text{if} ~ \left\|\mathbf{w}_{k}\right\|_2^{2} = 0 \\
	 1, \quad \text{otherwise} 
	 \end{array}
  \right. . 
\end{equation}
In this manner, the user scheduling, BS clustering and beamforming design for the downlink C-RAN
is unified within this single task of determining the sparse beamforming vector $\mathbf{w}_{k}$ 
for each user.

In this paper, we assume that the CP has access to global CSI
for designing the sparse beamforming vector $\mathbf{w}_{k}$.
Once $\mathbf{w}_{k}$ is determined, the CP transmits user $k$'s message, 
along with the beamforming coefficients, to
those BSs corresponding to the nonzero entries in $\mathbf{w}_{k}$ through
the backhaul links. 
We also assume that the channels are slow varying and only consider
the backhaul consumption due to the user data sharing and ignore the 
backhaul required for sharing CSI and delivering
beamforming coefficients.

Intuitively, the backhaul consumption at the $l$th BS is the accumulated data rates of the users
served by BS $l$. Notationally, we can characterize whether or not user $k$ is served by BS $l$ using the indicator function 
$\mathbbm{1}\left\{ \left\|\mathbf{w}_{k}^l\right\|_2^{2} \right\}$ 
and cast the per-BS backhaul constraint as:
\begin{equation}\label{eq:ExactBk}
\sum_{k \in \mathcal{K}} \mathbbm{1}\left\{ \left\|\mathbf{w}_{k}^l\right\|_2^{2} \right\} R_k \leq C_l, \quad \forall l  .
\end{equation}
Specifically, in the case where the serving cluster for each user is fixed, or equivalently the
set of users associated with each BS is predetermined, the backhaul consumption at BS $l$ 
is also equal to the accumulated data rates of users associated with BS $l$, which 
can be formulated as   
\begin{equation}
\sum_{k \in \mathcal{K}_l}  R_k \leq C_l, ~\forall l  \label{eq:ExactBk2}
\end{equation}
where $\mathcal{K}_l \subseteq \mathcal{K}$ denotes the fixed subset of users associated with BS $l$. 
Note that in each time-frequency slot, only the subset of users scheduled to be served have nonzero rates.
So in (\ref{eq:ExactBk2}), summing over the set of users associated with BS $l$ is equivalent to 
summing over the set of scheduled users.

From (\ref{eq:ExactBk}) and (\ref{eq:ExactBk2}), we see that the 
backhaul consumption is a function of both the cluster size and the user rate, where in addition 
the user rate is a function of user scheduling and beamforming operation. 
This observation provides us with different degrees of freedom in controlling the backhaul consumption depending on whether the BS  clustering is dynamic or fixed in different user scheduling time slots.
When the BS clustering is dynamic, we can jointly design the clustering, scheduling and beamforming to satisfy the per-BS backhaul constraint expressed in (\ref{eq:ExactBk}).
But even when the BS clustering is fixed (or static), we can still control the user rates through scheduling and beamforming vectors to make sure that the backhaul constraint expressed in (\ref{eq:ExactBk2}) is satisfied. 
In the following two sections, we discuss in detail how to incorporate the per-BS backhaul constraints in network utility maximization framework for downlink C-RAN under the above two different situations.

\section{Utility Maximization with Dynamic BS Clustering}

In this section, we propose a joint dynamic clustering, user scheduling and beamforming design strategy for the downlink C-RAN. 
The proposed algorithm designs a group sparse beamforming vector $\mathbf{w}_k$ for each user in each scheduling slot using 
an $\ell_1$-norm reweighting technique followed by a WMMSE approach under explicit per-BS backhaul capacity constraints.

\subsection{Problem Formulation}
This paper considers network utility maximization as the objective. 
Among the family of utility functions, WSR has been 
widely applied to network control and optimization problems.
In this paper, we also adopt the WSR utility but point out that the proposed scheme can be readily extend to any utility function that holds an equivalence relationship 
with the WMMSE problem (see \cite{Tom11} for a sufficient condition on the utility functions holding such an equivalence).

In dynamic BS clustering, the serving cluster for each user is a variable to be optimized in each scheduling slot.
We adopt the per-BS backhaul constraint formulation (\ref{eq:ExactBk}) and formulate
the WSR maximization problem under per-BS power constraints and per-BS backhaul constraints as:
\begin{subequations} \label{WSRwithBkhaul}
\begin{align} 
\maxi_{\left\{\mathbf{w}_k^l | l\in\mathcal{L}, k\in\mathcal{K}\right\}} \quad & \sum_{k\in\mathcal{K}} \alpha_k R_k \label{eq:obj1} \\ 
\sbto \quad &  \sum_{k\in\mathcal{K}} \left\|\mathbf{w}_{k}^l\right\|_2^{2} \leq P_l, ~\forall l \\
& \sum_{k\in\mathcal{K}} \mathbbm{1}\left\{ \left\|\mathbf{w}_{k}^l\right\|_2^{2} \right\} R_k \leq C_l, ~\forall l  \label{bkhaulconst}
\end{align}
\end{subequations}
%\sum_{k} \left\| \|\mathbf{w}_{k}^l\|_2^{2} \right\|_0 R_k \leq C_l, ~\forall l  \label{bkhaulconst}
where $\alpha_k$ denotes the priority weight associated with user $k$ at the current scheduling slot which can be updated
according to, for example, the proportional fairness criterion.
Here, $P_l$ and $C_l$ represent
the transmit power budget and the backhaul capacity limit for BS $l$, respectively.
The rate $R_k$ shown in (\ref{eq:obj1}) and (\ref{bkhaulconst}) is defined in (\ref{eq:Rk2}), which is a function of the set of 
sparse beamforming vectors $\mathbf{w}_k$'s only. 
Note that $\mathbf{w}_k$ is comprised of the beamforming vectors $\mathbf{w}_{k}^l$'s.
Thus, the optimization variables for 
problem (\ref{WSRwithBkhaul}) are the set of $\mathbf{w}_{k}^l$'s.

\subsection{Proposed Algorithm}
The conventional WSR maximization problem 
is a well-known nonconvex optimization problem, for which finding the global optimal solution 
is already quite challenging even without the additional mixed discrete and continuous 
backhaul constraint (\ref{bkhaulconst}). 
This paper focuses on heuristic algorithms for approaching a local optimum 
solution to the problem (\ref{WSRwithBkhaul}) only. 
Our main contribution is a new way of dealing with the discrete indicator function in constraint (\ref{bkhaulconst}).

A key observation made in this paper is that the indicator function in (\ref{bkhaulconst}) can also be equivalently expressed as an 
$\ell_0$-norm of a scalar. 
The $\ell_0$-norm is the number of nonzero entries in a vector. So it reduces to an indicator function in the scalar case. 
This equivalent expression allows us to use ideas from the 
compressive sensing literature \cite{Candes08}, where a nonconvex $\ell_0$-norm optimization objective can often be 
approximated by a convex reweighted $\ell_1$-norm, i.e. 
\begin{align} \label{eq:l1Rewht}
\left\| \mathbf{x} \right\|_0 \approx  \sum_i \beta_i |x_i|
\end{align}
where $x_i$ denotes the $i$th component in the vector $\mathbf{x}$ and $\beta_i$ is the weight associated with $x_i$.
By properly choosing weights $\beta_i$'s, the minimization of $\left\| \mathbf{x} \right\|_0$ can be effectively solved through the 
minimization of $\sum_i \beta_i |x_i|$ instead.

This paper goes one step further in that we extend the reweighted $\ell_1$-norm approximation technique (\ref{eq:l1Rewht}) originally proposed for minimizing the $\ell_0$-norm in the objective to dealing with the $\ell_0$-norm in the constraint.
In particular, we rewrite the indicator function $\mathbbm{1}\left\{ \left\|\mathbf{w}_{k}^l\right\|_2^{2} \right\}$ as 
\begin{equation}
\mathbbm{1}\left\{ \left\|\mathbf{w}_{k}^l\right\|_2^{2} \right\} = \left\| \left\|\mathbf{w}_{k}^l\right\|_2^{2} \right\|_0 ,
\end{equation}
and reformulate the backhaul constraint (\ref{bkhaulconst}) as:
\begin{equation} \label{approxbkhaulconst}
\sum_{k\in\mathcal{K}}  \beta_k^{l} \left\|\mathbf{w}_{k}^l\right\|_2^{2} R_k \leq C_l 
\end{equation}
where $\beta_k^{l}$ is a constant weight associated with BS $l$ and user $k$ and is updated iteratively according to 
\begin{equation} \label{UpdateBeta}
\beta_k^{l} = \frac{1}{\left\|\mathbf{w}_{k}^l\right\|_2^{2} + \tau}, \forall k, l 
\end{equation}
with some small constant regularization factor $\tau > 0$ and 
$\left\|\mathbf{w}_{k}^l\right\|_2^{2}$ from the previous iteration. 

The heuristic weight updating rule (\ref{UpdateBeta}) is motivated by the fact 
that by choosing $\beta_k^{l}$ to be inversely proportional to the transmit power level $\left\|\mathbf{w}_{k}^l\right\|_2^{2}$, those
BSs with lower transmit power to user $k$ would have higher weights and would be forced to further reduce its transmit power 
and encouraged to drop out of the BS cluster eventually. 
Note that not only the BS cluster formation, but also the user scheduling can be controlled 
through $\left\|\mathbf{w}_{k}^l\right\|_2^{2}$ since 
the user $k$ is scheduled if and only if there exists at least one BS $l \in \mathcal{L}$ such that $\left\|\mathbf{w}_{k}^l\right\|_2^{2} \neq 0$.

However, even with the above approximation, the optimization problem (\ref{WSRwithBkhaul})
with the backhaul constraint (\ref{bkhaulconst}) replaced by 
(\ref{approxbkhaulconst}) is still difficult to deal with, due to the 
fact that the rate $R_k$ appears in both the objective function and the constraints. 
To address this difficulty, we propose to solve the problem (\ref{WSRwithBkhaul}) iteratively with fixed rate
$\hat{R}_k$ in (\ref{approxbkhaulconst}) obtained from the previous iteration. 
Under fixed $\beta_k^{l}$ and $\hat{R}_k$, problem (\ref{WSRwithBkhaul}) now reduces to% the following optimization problem: 
\begin{subequations} \label{AppWSRwithBkhaul}
\begin{align} 
\maxi_{\left\{\mathbf{w}_k^l| l\in\mathcal{L}, k\in\mathcal{K}\right\}} \quad & \sum_{k\in\mathcal{K}} \alpha_k R_k  \\
\sbto \quad &  \sum_{k\in\mathcal{K}} \left\|\mathbf{w}_{k}^l\right\|_2^{2} \leq P_l, ~\forall l  \label{finalpowerconst} \\
& \sum_{k\in\mathcal{K}} \beta_k^{l} \hat{R}_k \left\|\mathbf{w}_{k}^l\right\|_2^{2} \leq C_l, ~\forall l  \label{finalbkhaulconst}
\end{align}
\end{subequations}
where the approximated backhaul constraint (\ref{finalbkhaulconst}) can be interpreted as a \emph{weighted}
per-BS power constraint bearing a resemblance to the traditional per-BS power constraint (\ref{finalpowerconst}).

Although the approximated problem (\ref{AppWSRwithBkhaul}) is still nonconvex, 
we can reformulate it as an 
equivalent WMMSE problem and use the block coordinate descent method 
to reach a stationary point of (\ref{AppWSRwithBkhaul}). 
The equivalence between WSR maximization and WMMSE 
is first established in \cite{Cioffi08} for MIMO broadcast channel 
and later generalized to MIMO interfering channel in \cite{Tom11} and 
MIMO interfering channel with partial cooperation in \cite{Kaviani12}.

It is not difficult to see that the generalized WMMSE equivalence
established in \cite{Tom11} also extends to the problem (\ref{AppWSRwithBkhaul})
with the newly introduced weighted per-BS power constraint (\ref{finalbkhaulconst}).
We explicitly state the equivalence as follows:
\begin{proposition}
\label{prop:WMMSE}
The WSR maximization problem (\ref{AppWSRwithBkhaul}) has the same optimal solution as the
following WMMSE problem:
\begin{align} \label{WMSE}
\mini_{\left\{\rho_k, \mathbf{u}_{k}, \mathbf{w}_k^{l}| l\in\mathcal{L}, k\in\mathcal{K}\right\}} \quad & \sum_{k\in\mathcal{K}} \alpha_k \left(\rho_k e_k - \log \rho_k \right)   \\ \nonumber
\sbto \quad &  \sum_{k\in\mathcal{K}} \left\|\mathbf{w}_{k}^l\right\|_2^{2} \leq P_l, ~\forall l    \\ \nonumber 
& \sum_{k\in\mathcal{K}} \beta_k^{l} \hat{R}_k \left\|\mathbf{w}_{k}^l\right\|_2^{2} \leq C_l, ~\forall l  \nonumber
\end{align}
where $\rho_k$ denotes the MSE weight for user $k$ and $e_k$ is the corresponding MSE defined as
\begin{align} 
 e_k  = ~&\mathsf{E} \left [\left\| \mathbf{u}_{k}^{H} \mathbf{y}_k - s_k \right\|_{2}^2 \right ]   \nonumber \\
  = ~& \mathbf{u}_{k}^{H} \left ( \sum_{j\in\mathcal{K}} \mathbf{H}_{k} \mathbf{w}_j \mathbf{w}_{j}^{H} \mathbf{H}_{k}^{H} + \sigma^2 \mathbf{I}  \right ) \mathbf{u}_{k}   \nonumber \\
&  - 2 \text{Re} \left\{ \mathbf{u}_{k}^{H} \mathbf{H}_{k} \mathbf{w}_k\right\} + 1 \label{eq:MSE}
\end{align}
under the receiver $\mathbf{u}_{k} \in \mathbb{C}^{N \times 1}$. 
\end{proposition}
%\begin{proof}
%We omit the proof due to limited space. 
%\end{proof}

The advantage of solving WSR maximization problem (\ref{AppWSRwithBkhaul}) 
through its equivalent WMMSE problem (\ref{WMSE}) is that
(\ref{WMSE}) is convex with respect to each of the individual optimization variables.
This crucial fact allows problem (\ref{WMSE}) to be solved efficiently
through the block coordinate descent method 
by iteratively optimizing over $\rho_k$, $\mathbf{u}_{k}$ and $\mathbf{w}_{k}$:
\begin{itemize}
\item The optimal MSE weight $\rho_k$ under fixed $\mathbf{w}_{k}$ and $\mathbf{u}_{k}$
is given by 
\begin{align} \label{eq:rho}
\rho_k = e_k^{-1}, \quad \forall k .
\end{align} 
\item The optimal receiver $\mathbf{u}_{k}$ under fixed $\mathbf{w}_{k}$ and $\rho_k$ is the 
MMSE receiver:
\begin{equation}
\mathbf{u}_k = \left (\sum_{j\in\mathcal{K}} \mathbf{H}_{k} \mathbf{w}_j \mathbf{w}_{j}^{H} \mathbf{H}_{k}^{H} + \sigma^2 \mathbf{I} \right )^{-1} \mathbf{H}_{k} \mathbf{w}_k , \quad \forall k .
\label{eq:MSEreceiver}
\end{equation}
\item The optimization problem for finding the optimal transmit beamformer $\mathbf{w}_{k}$ 
under fixed $\mathbf{u}_{k}$ and $\rho_k$ is a quadratically constrained 
quadratic programming (QCQP) problem: 
\begin{align}\label{Prob:QCQP}
\mini_{\left\{\mathbf{w}_k^{l}| l\in\mathcal{L}, k\in\mathcal{K}\right\}} \quad & \sum_{k\in\mathcal{K}}  
\mathbf{w}_{k}^{H} \left ( \sum_{j\in\mathcal{K}} \alpha_j \rho_j \mathbf{H}_{j}^{H} \mathbf{u}_j \mathbf{u}_{j}^{H} \mathbf{H}_{j} \right ) \mathbf{w}_{k} \\  \nonumber
& \quad - 2 \sum_{k\in\mathcal{K}} \alpha_k \rho_k \text{Re} \left\{ \mathbf{u}_{k}^{H} \mathbf{H}_{k} \mathbf{w}_k\right\}   \\ \nonumber
\sbto \quad &  \sum_{k\in\mathcal{K}} \left\|\mathbf{w}_{k}^l\right\|_2^{2} \leq P_l, ~\forall l    \\ \nonumber 
& \sum_{k\in\mathcal{K}} \beta_k^{l} \hat{R}_k \left\|\mathbf{w}_{k}^l\right\|_2^{2} \leq C_l, ~\forall l  \nonumber
\end{align}
which can be solved using a standard convex optimization solver such as CVX \cite{cvx}.
\end{itemize}

A straightforward way of applying the above WMMSE algorithm to solve the original
problem (\ref{WSRwithBkhaul}) would involve two loops:
an inner loop
to solve the approximated WSR maximization problem (\ref{AppWSRwithBkhaul}) 
with fixed weight $\beta_k^{l}$ and rate $\hat{R}_k$,
and an outer loop to update $\beta_k^{l}$ and $\hat{R}_k$.
Although such an algorithm can guarantee that the inner loop converges to a stationary point of the problem
(\ref{AppWSRwithBkhaul}), its computational complexity can be high. 
Instead, we propose to combine these two loops into a single loop and
update the weight $\beta_k^{l}$ and rate $\hat{R}_k$
inside the WMMSE algorithm, as summarized in Algorithm~\ref{alg:LCSparse}.
Although Algorithm~\ref{alg:LCSparse} does not have a proof of convergence, numerical simulation shows that 
it converges reasonably fast, and it significantly outperforms the fixed clustering baseline schemes.

\begin{algorithm}
{\bf Initialization}: $\beta_{k}^{l}, \hat{R}_k, \mathbf{w}_{k},\forall l, k $; \\
{\bf Repeat}:
\begin{enumerate}
\item Fix $\mathbf{w}_{k}, \forall k$, compute the MMSE receiver $\mathbf{u}_{k}$ and the corresponding MSE $e_k$ according to (\ref{eq:MSEreceiver}) and (\ref{eq:MSE});
\item Update the MSE weight $\rho_k$ according to (\ref{eq:rho});
\item Find the optimal transmit beamformer $\mathbf{w}_{k}$ under fixed $\mathbf{u}_{k}$ and $\rho_k, \forall k$, 
by solving the QCQP problem (\ref{Prob:QCQP}); 
\item Compute the achievable rate $R_k$ according to (\ref{eq:Rk2}), $\forall k$;
\item Update $\hat{R}_k = R_k$ and $\beta_{k}^{l}$ according to (\ref{UpdateBeta}), $\forall l, k$.
\end{enumerate}
{\bf Until} convergence
\caption{WSR Maximization with Per-BS Backhaul Constraints under Dynamic BS Clustering}
\label{alg:LCSparse}
\end{algorithm}

\subsection{Complexity Analysis}

Assuming a typical network with $K > L > M > N$, the computational complexity of Step 1 in Algorithm~\ref{alg:LCSparse} 
is $O(K^2LMN)$, mainly due to the receive covariance matrix computation 
in (\ref{eq:MSEreceiver}) and (\ref{eq:MSE}).  
With the MSE $e_k$ obtained from Step 1, the additional computational complexity for Step 2 for updating all the MSE weights $\rho_k$'s 
is only $O(K)$. 
Step 3 requires solving a QCQP problem, which can also be equivalently
reformulated as a second order cone programming (SOCP) problem as we do in the simulation part of this paper. 
The total number of variables in the equivalent SOCP problem is $KLM$ and the computation complexity of using 
interior-point method to solve such an SOCP problem is approximately $O((KLM)^{3.5})$ \cite{Ye98}.
In Step 5, the rate updating procedure requires the computation of the achievable rate in Step 4 according to (\ref{eq:Rk2}), 
which has the same computational complexity 
as computing the MSE, i.e. $O(K^2LMN)$. %, while the weight updating rule (\ref{UpdateBeta}) requires 
%$\mathcal{O}(KLN^2)$ computational burden to update all the weights $\beta_{k}^{l}$'s. 
As we can see, the computational complexity of Algorithm~\ref{alg:LCSparse} per iteration mainly
comes from the optimal transmit beamformer design in Step 3.
Suppose Algorithm~\ref{alg:LCSparse} requires $T$ total number of iterations to converge, the overall 
computational complexity of Algorithm~\ref{alg:LCSparse} is therefore $O((KLM)^{3.5}T)$.

\subsection{Heuristic Complexity Reduction Techniques}

To improve the efficiency of Algorithm~\ref{alg:LCSparse} in each iteration, 
in what follows, we further propose two techniques, \emph{iterative link removal} and \emph{iterative user pool shrinking}.
The former aims at reducing the number of potential transmit antennas $LM$ serving each user while the latter is intended to decrease the total number of users $K$ to be considered in each iteration.

\subsubsection{Iterative Link Removal}

Similar to what we observed in \cite{binbin13}, the transmit power from some of the candidate serving BSs 
would drop down rapidly as the iterations go on.
By taking advantage of this fact, we propose to iteratively remove the $l$th BS from the $k$th user's candidate cluster once the transmit power from BS $l$ to user $k$, i.e. $\left\|\mathbf{w}_{k}^l\right\|_2^{2}$, is below a certain threshold, say $-100$ dBm/Hz.
This reduces the dimension of the potential transmit beamformer for each user and reduces the 
complexity of solving SOCP in Step 3 of Algorithm~\ref{alg:LCSparse}.

\subsubsection{Iterative User Pool Shrinking}

The proposed algorithm does user scheduling implicitly.
We observe from simulations that, it is beneficial for Algorithm~\ref{alg:LCSparse}
to consider a large pool of users. 
However, to consider all the users in the entire network all the time would incur
significant computational burden. 
Instead, we propose to check the achievable user rate $R_k$ 
in Step 4 in each iteration and ignore those users with 
negligible rates (below some threshold, say 0.01 bps/Hz) for subsequent iterations. 
Our simulations show that, after around 10 iterations, more than half of the total
users can be taken out of the consideration 
with negligible performance loss to the overall algorithm. This significantly
reduces the total number of variables to be optimized.
%and the efficiency of the proposed algorithm is significantly improved. 
%To demonstrate the effectiveness of this technique, we plot the total number of survived users whose rates are above the threshold at each iteration in Fig.~\ref{fig:NumSurviveUsers}, from which we can see that after 7 iterations the number of users to be potentially scheduled can be reduced to half of the initial entire user pool. 

\section{Utility Maximization with Static BS Clustering}

In the previous section, BS clustering is dynamically determined 
in each time-frequency slot together with the beamforming vector and user scheduling in a joint fashion. 
However, dynamic BS clustering may incur significant signaling overhead in practice as new BS-user associations 
need to be established continuously over time. 
In this section, we discuss static clustering schemes, where the BS clusters only need to be updated at much 
larger time scale, typically only when user locations change.

As discussed previously, backhaul consumption under static clustering can still be controlled  
by jointly optimizing the user scheduling and beamforming. 
In this section, we first adapt the sparse beamforming algorithm proposed in previous section to jointly schedule the users and design the beamforming vectors under per-BS backhaul constraints while assuming that 
the BS clustering is fixed. 
We then propose two heuristic static clustering schemes: one depends on the maximum number of users each BS can support and the
long-term channel condition each user experiences; the other generalizes the SINR-bias technique used for 
cell range expansion \cite{Guvenc11} to form a static cluster for each user.

\subsection{Joint Scheduling and Beamforming Design with Fixed BS Clustering}

\begin{figure*}[!t]
% ensure that we have normalsize text
\normalsize
% Store the current equation number.
\setcounter{MYtempeqncnt}{\value{equation}}
% Set the equation number to one less than the one
% desired for the first equation here.
% The value here will have to changed if equations
% are added or removed prior to the place these
% equations are referenced in the main text.
\setcounter{equation}{20}
\begin{equation}
\label{eqn_dbl_rx}
\mathbf{u}_k = \left (\sum_{j\in\mathcal{K}} \mathbf{H}_{k}^{\mathcal{L}_j} \mathbf{w}_j^{\mathcal{L}_j} \left (  \mathbf{H}_{k}^{\mathcal{L}_j} \mathbf{w}_j^{\mathcal{L}_j}  \right)^{H} + \sigma^2 \mathbf{I} \right )^{-1} \mathbf{H}_{k}^{\mathcal{L}_k} \mathbf{w}_k^{\mathcal{L}_k}
\end{equation}
\begin{equation}
\label{eqn_dbl_mse}
 e_k  = \mathbf{u}_{k}^{H} \left (\sum_{j\in\mathcal{K}} \mathbf{H}_{k}^{\mathcal{L}_j} \mathbf{w}_j^{\mathcal{L}_j} \left (  \mathbf{H}_{k}^{\mathcal{L}_j} \mathbf{w}_j^{\mathcal{L}_j}  \right)^{H} + \sigma^2 \mathbf{I} \right ) \mathbf{u}_{k}  - 2 \text{Re} \left\{ \mathbf{u}_{k}^{H} \mathbf{H}_{k}^{\mathcal{L}_k} \mathbf{w}_k^{\mathcal{L}_k}\right\} + 1
\end{equation}
\begin{equation}
\label{eqn_dbl_rate}
R_k = \log \left( 1 +\left (  \mathbf{H}_{k}^{\mathcal{L}_k} \mathbf{w}_k^{\mathcal{L}_k}  \right)^{H} \left (\sum_{j \neq k, j \in\mathcal{K}} \mathbf{H}_{k}^{\mathcal{L}_j} \mathbf{w}_j^{\mathcal{L}_j} \left (  \mathbf{H}_{k}^{\mathcal{L}_j} \mathbf{w}_j^{\mathcal{L}_j}  \right)^{H} + \sigma^2 \mathbf{I} \right )^{-1} \mathbf{H}_{k}^{\mathcal{L}_k} \mathbf{w}_k^{\mathcal{L}_k} \right) 
\end{equation}
% Restore the current equation number.
\setcounter{equation}{\value{MYtempeqncnt}}
% The spacer can be tweaked to stop underfull vboxes.
\vspace*{4pt}
% IEEE uses as a separator
\hrulefill
\end{figure*}

Let $\mathcal{L}_k$ be the fixed cluster of BSs serving user $k$. 
The joint scheduling and beamforming design problem is that of determining the scheduled users in each time-frequency slot and
the corresponding beamformers from the BSs in $\mathcal{L}_k$ to each scheduled user $k$ 
while satisfying the per-BS power constraints and per-BS backhaul constraints.

Equivalently, let $\mathcal{K}_l$ be the set of users associated with BS $l$, the network utility maximization problem can now 
be formulated as 
\begin{subequations} \label{WSRwithFixBS}
\begin{align} 
\maxi_{\left\{\mathbf{w}_k^l| k\in\mathcal{K}, l \in \mathcal{L}_k \right\}} \quad & \sum_{k\in\mathcal{K}} \alpha_k R_k \\
\sbto \quad &  \sum_{k \in \mathcal{K}_l} \left\|\mathbf{w}_{k}^l\right\|_2^{2} \leq P_l, ~\forall l  \label{PerBSPowerFixBS} \\
& \sum_{k \in \mathcal{K}_l}  R_k \leq C_l, ~\forall l  \label{PerBSBkhaulFixBS}
\end{align}
\end{subequations}
Note that the difference between the utility maximization problems (\ref{WSRwithFixBS}) and (\ref{WSRwithBkhaul}) is that the
transmit power and backhaul constraint for BS $l$ now only need to 
take into account the fixed subset of users associated with BS $l$, $\mathcal{K}_l$. 
Also, the optimization variable $\mathbf{w}_k^l$ is
only over the beamforming vectors from the BSs in each user $k$'s serving cluster $\mathcal{L}_k$ since $\mathbf{w}_k^l = \mathbf{0}$ for 
$l \notin \mathcal{L}_k$.

Note that user scheduling is implicitly being optimized in (\ref{WSRwithFixBS}). 
Only the subset of users scheduled in the current time-frequency slot would have nonzero rates and 
need to be considered in the summations (\ref{PerBSPowerFixBS}) and (\ref{PerBSBkhaulFixBS}). 
With this observation, we can rewrite 
(\ref{PerBSBkhaulFixBS}) in the following equivalent form:
\begin{align} \label{PerBSBkhaulFixBS2}
\sum_{k \in \mathcal{K}_l} \mathbbm{1}\left\{ \left\|\mathbf{w}^{\mathcal{L}_k}_k \right\|_2^2 \right\}  R_k \leq C_l, ~\forall l .
\end{align}
where $\mathbf{w}^{\mathcal{L}_k}_k \in \mathbb{C}^{|\mathcal{L}_k |M \times 1}$ is the beamforming vector from user $k$'s serving 
cluster $\mathcal{L}_k$ to user $k$. 
This allows us to utilize a similar 
idea as in previous section to solve problem (\ref{WSRwithFixBS}) approximately by 
fixing the user rate $R_k$ in the constraint and approximating the indicator function in 
(\ref{PerBSBkhaulFixBS2}) using the reweighted $\ell_1$-norm technique.  
The resulting approximated optimization problem to (\ref{WSRwithFixBS}) now becomes
\begin{subequations} \label{AppWSRwithBkhaul2}
\begin{align} 
\maxi_{\left\{\mathbf{w}_k^l| k\in\mathcal{K}, l \in \mathcal{L}_k \right\}} \quad & \sum_{k\in\mathcal{K}} \alpha_k R_k  \\
\sbto \quad &  \sum_{k \in \mathcal{K}_l} \left\|\mathbf{w}_{k}^l\right\|_2^{2} \leq P_l, ~\forall l  \\
& \sum_{k \in \mathcal{K}_l} \beta_k \hat{R}_k \left\|\mathbf{w}^{\mathcal{L}_k}_k \right\|_2^2 \leq C_l, ~\forall l  \label{finalbkhaulconst2}
\end{align}
\end{subequations}
where $\beta_k$ is the constant weight associated with the predetermined BS cluster $\mathcal{L}_k$ for user $k$ and is updated
iteratively according to
\begin{equation} \label{UpdateBetak}
\beta_k = \frac{1}{\left\|\mathbf{w}^{\mathcal{L}_k}_k \right\|_2^2 + \tau}, \forall k ,
\end{equation}
and $\hat{R}_k$ is the achievable rate from previous iteration.

Comparing the weight updating rule (\ref{UpdateBetak}) with (\ref{UpdateBeta}), we note that the role of $\beta_k^l$ in
problem formulation (\ref{AppWSRwithBkhaul}) is to determine whether or not BS $l$ should serve user $k$ in the current 
time-frequency slot, while the role of $\beta_k$ in problem formulation (\ref{AppWSRwithBkhaul2}) is to decide whether or not user $k$ 
should be scheduled and served by
its predetermined serving cluster $\mathcal{L}_k$ \emph{as a whole} in the current time-frequency slot. 
Note that $\beta_k^l$ only appears in the $l$th BS backhaul constraint in (\ref{finalbkhaulconst}), while 
in (\ref{finalbkhaulconst2}) $\beta_k$ appears in the approximated per-BS backhaul 
constraints corresponding to all of the user $k$'s pre-associated cluster of BSs in $\mathcal{L}_k$.

With the approximated problem formulation (\ref{AppWSRwithBkhaul2}), it becomes straightforward to extend Algorithm~\ref{alg:LCSparse} 
for solving problem (\ref{WSRwithFixBS}). The only necessary adaptation occurs in Step 5 of Algorithm~\ref{alg:LCSparse}, where the update of weight $\beta_k^l$ is replaced by updating the weight $\beta_k$ according to (\ref{UpdateBetak}).
Under given BS cluster $\mathcal{L}_k$, we rewrite the 
MMSE receiver, MSE and achievable rate for user $k$ in equation
(\ref{eqn_dbl_rx}), (\ref{eqn_dbl_mse}) and  (\ref{eqn_dbl_rate}) respectively\footnote{One can also use the MMSE receiver, MSE and achievable rate defined previously in (\ref{eq:MSEreceiver}), (\ref{eq:MSE}) and (\ref{eq:Rk2}) respectively by filling those entries corresponding 
to $\mathcal{L}_k$ in $\mathbf{w}_k$ with $\mathbf{w}^{\mathcal{L}_k}_k$ and the rest entries in $\mathbf{w}_k$ with zero.}, where $\mathbf{H}_{k}^{\mathcal{L}_j}$ is
the CSI matrix from user $j$'s serving cluster $\mathcal{L}_j$ to user $k$. 
The proposed joint scheduling and beamforming design algorithm under fixed BS clustering 
is summarized in Algorithm~\ref{alg:LCSparseStatic}.

\begin{algorithm}
{\bf Initialization}: $\beta_{k}, \hat{R}_k, \mathbf{w}^{\mathcal{L}_k}_k,\forall k $; \\
{\bf Repeat}:
\begin{enumerate}
\item Fix $\mathbf{w}^{\mathcal{L}_k}_k, \forall k$, compute the MMSE receiver $\mathbf{u}_{k}$ and the corresponding MSE $e_k$ according to (\ref{eqn_dbl_rx}) and (\ref{eqn_dbl_mse});
\item Update the MSE weight $\rho_k$ according to (\ref{eq:rho});
\item Find the optimal transmit beamformer $\mathbf{w}^{\mathcal{L}_k}_k$ under fixed $\mathbf{u}_{k}$ and $\rho_k, \forall k$, by solving the QCQP problem (\ref{Prob:QCQP}) with $\mathbf{w}_k$, $\mathbf{H}_j$ and $\mathbf{H}_k$ replaced by 
$\mathbf{w}^{\mathcal{L}_k}_k$, $\mathbf{H}_{j}^{\mathcal{L}_k}$ and $\mathbf{H}_{k}^{\mathcal{L}_k}$ respectively; 
\item Compute the achievable rate $R_k$ according to (\ref{eqn_dbl_rate}), $\forall k$;
\item Update $\hat{R}_k = R_k$ and $\beta_{k}$ according to (\ref{UpdateBetak}), $\forall k$.
\end{enumerate}
{\bf Until} convergence
\caption{WSR Maximization with Per-BS Backhaul Constraints under Static BS Clustering}
\label{alg:LCSparseStatic}
\end{algorithm}

It is worth noting that both Algorithm~\ref{alg:LCSparse} and~\ref{alg:LCSparseStatic} implement user scheduling 
operation implicitly by optimizing the beamforming vectors for all the users in the entire network but only selecting those users with nonzero beamforming vectors to be served. 
This is in contrast to the conventional user scheduling approach, 
where typically a subset of users are pre-selected and only the beamforming vectors
corresponding to those pre-selected users are optimized. 
Simulation results show that the proposed algorithms are able to achieve better 
performance by scheduling the users implicitly, although the performance gain comes at a complexity cost. 

We also note that both Algorithm~\ref{alg:LCSparse} and~\ref{alg:LCSparseStatic} require global CSI at the CP in order to schedule
the users and to design beamformers accordingly, which may lead to large channel estimation overhead. 
Performance analysis of C-RAN with partial CSI has been carried on in \cite{Mohiuddin13} under a simplified model where the BSs 
and the users are equipped with a single antenna each and no backhaul constraint is considered. 
The impact of channel estimation overhead on the system performance of C-RAN under a more realistic model considered 
in this paper is nontrivial and is left for future work.

\subsection{Proposed Static Clustering Algorithms}

Thus far in this section, we have dealt with the joint user scheduling and beamforming design under per-BS backhaul capacity constraints 
for any given fixed clustering scheme. 
We now propose heuristic algorithms to optimize over the clustering strategies.

Differing from the traditional BS-user association problem where each user is only associated with a single BS, 
in C-RAN each user is served by a cluster of BSs. 
The optimal fixed BS clustering design for C-RAN is nontrivial as each user wants to be served by
as many nearby BSs as possible while each BS can only support a limited number of users due to the limited
radio resources, i.e. transmit power, and limited backhaul capacity.
A good clustering strategy for C-RAN should account for not only the channel strength from the BSs to 
each user but also the available resources at each BS.

This paper adopts the user-centric clustering strategy, in which each user is served by an individually selected and potentially overlapping subset of BSs.
A simple way of forming user-centric clusters is to choose an equal number of strongest BSs around each user
as its serving cluster. 
However, such a scheme may produce imbalanced traffic loads across the network, especially in a heterogeneous
deployment where the macro-BSs typically have much higher transmit power than the pico-BSs, 
so more than the optimal number of users would associate with the macro-BSs.
This paper proposes two heuristic static clustering schemes to address this load balancing issue. The first scheme is based on imposing a maximum load on each BS. The second scheme is based on introducing a bias term to the received signal strength at each user. These two schemes are described in detail below.

\subsubsection{Maximum Loading Based Static Clustering}

In order to avoid BS overloading, in this scheme, we propose to set an upper bound on the number of users that can associate with
each BS $l$, denoted as $K_{l,max}$. The value of $K_{l,max}$ depends on the amount of resources available at BS $l$. 
For example, in a heterogeneous network, the macro-BSs usually have higher transmit power and more backhaul capacity than the pico-BSs, 
hence $K_{l,max}$ for macro-BSs should be larger than that of pico-BSs. With $K_{l,max}$, those BSs that have already reached its maximum number of associated users would direct the subsequent users to other underloaded BSs.

At the user's side, each user wants to be served by as many nearby BSs as possible to obtain the highest service rate. 
However, since a cell-center user typically already 
experiences good channel condition, it is reasonable to connect it with fewer BSs, whereas a cell-edge user may need
more serving BSs to coordinately mitigate inter-cell interference. 
To capture the difference in the ideal cluster size for different users, which depends on their relative locations, 
we propose to set a candidate BS cluster for each user based on a threshold on the received signal strength difference. 
For each user $k$, only those BSs from which the received signal strength is within $\eta_1$ gap to the signal strength from the strongest BS 
are considered as potential serving BSs. 
Mathematically, let $s_{l,k}$ be the received signal strength from BS $l$ to user $k$, defined as the maximum transmit 
power from BS $l$ compensated by the path loss to user $k$ without accounting for possible antenna beamforming gain, 
the candidate serving cluster $\mathcal{C}_k$ for user $k$ is a set 
defined as follows:
\setcounter{equation}{23}
\begin{align} \label{eq:CandidateCluster}
\mathcal{C}_k = \left\{l \in\mathcal{L} \vert ~ \max_{m} s_{m,k} - s_{l,k} \leq \eta_1 \right\} .
\end{align}
Since a cell-edge user sees more nearby BSs with similar signal strength than a cell-center user, 
the candidate cluster size $|\mathcal{C}_k|$ for a 
cell-edge user would be larger, which potentially results in more serving BSs for the cell-edge user.

Based on the aforementioned parameter $K_{l,max}$ and set $\mathcal{C}_k$, we propose
a simple user-centric clustering scheme based on the following two heuristics:
\begin{itemize}
\item Each user $k$ sends requests to the BSs in the candidate set $\mathcal{C}_k$ sequentially from the strongest to the weakest;
\item Each BS $l$ accepts up to $K_{l,max}$ users.
\end{itemize}
The details of the first proposed static clustering scheme is shown in Algorithm~\ref{alg:GMRCF}, which requires
a multi-round negotiation between the BSs and the users until the BSs are connected with the maximum
number of users or the users have exhausted all their BSs in the candidate sets.

\begin{algorithm}
{\bf Initialization}: 
\begin{enumerate}
\item $\mathcal{K} = \{1,2,\cdots,K\}$, $\mathcal{L} = \{1,2,\cdots,L\}$;
%\item User $k$'th serving cluster $\mathcal{L}_k = \emptyset, \forall k$;
\item Let $\mathcal{K}_l$ be the set of users associated with the $l$th BS. Set $\mathcal{K}_l = \emptyset, l=1,2, \cdots,L$;
\item Let $K_{l,max}$ be the maximum number of users BS $l$ can support, $l=1,2,\cdots,L$;
%\item Maximum number of BSs user $k$ can be served by: $L_{k,max}, \forall k$;
\item Let $\mathcal{C}_k$ be the candidate serving cluster for user $k$, $k=1,2,\cdots,K$;
\item Set iteration index $i=1$.
\end{enumerate}
{\bf Repeat}: 
\begin{enumerate}
\item Each user $k \in \mathcal{K}$ sends a request to the $i$th strongest BS in $\mathcal{C}_k$;

\item For each BS $l \in \mathcal{L}$:\\
		{\bf If} $K_{l,max} - |\mathcal{K}_l| \geq$ total number of received requests
			\begin{enumerate}
			\item[2.1)]  $\mathcal{K}_l = \mathcal{K}_l ~ \cup$ \{All the received requests\};
			\end{enumerate}
		{\bf otherwise}
			\begin{enumerate}
			\item[2.2)]	 $\mathcal{K}_l = \mathcal{K}_l ~ \cup $  
			\{The $\left(K_{l,max} - |\mathcal{K}_l|\right)$ strongest users among all the received requests\};
			\item[2.3)] $\mathcal{L} = \mathcal{L} \setminus \{l\}$.
			\end{enumerate}
		{\bf end if}

\item For each user $k \in \mathcal{K}$, if it has exhausted all the candidate BSs in the list $\mathcal{C}_k$, 
update $\mathcal{K} = \mathcal{K} \setminus \{k\}$;
			%{\bf If}  $i = |\mathcal{C}_k|$ 
		  %\begin{enumerate}
			%\item[4.1)]	Update $\mathcal{K} = \mathcal{K} \setminus \{k\}$;
			%\end{enumerate}
%\item[] {\bf end if}

\item $i = i + 1$.
\end{enumerate}
{\bf Until} $\mathcal{L} = \emptyset$ or $\mathcal{K} = \emptyset$
\caption{Maximum Loading Based Static Clustering}
\label{alg:GMRCF}
\end{algorithm}

We remark that the parameters $K_{l,max}$ and $\mathcal{C}_k$ jointly play an important role in 
determining the serving clusters in Algorithm~\ref{alg:GMRCF}. 
Using $K_{l,max}$ or $\mathcal{C}_k$ alone would not have produced a good clustering scheme. 
For instance, suppose each BS is simply associated with the $K_{l,max}$ strongest users it sees, then the 
cell-edge users would be at a risk of connecting with no BSs. 
Or, if each user is served by all the BSs in its candidate list $\mathcal{C}_k$, then the high-power BSs may be overloaded.
Only by taking into account both the traffic load of each BS and the channel condition of each user 
through the parameters $K_{l,max}$ and $\mathcal{C}_k$ jointly
would the proposed Algorithm~\ref{alg:GMRCF} be able to produce a good clustering scheme.

It is worth noting that the use of $(K_{l,max}$, $\mathcal{C}_k)$ in Algorithm~\ref{alg:GMRCF} is only one possibility in jointly 
controlling the traffic load of the BSs and the channel condition of the users. 
For example, one can also set the candidate cluster for user $k$ as the strongest $L_k$ BSs instead of the proposed
$\mathcal{C}_k$. Such an algorithm also has reasonable performance. 
However, to find a good $L_k$ for each user in the entire network is not 
an easy task. Suppose that the $L_k$ is set to be equal for all users to simplify the search, it may result in the cell-center users and the cell-edge users being associated with an equal (or at least a similar) number of BSs, 
which may lead to inefficient usage of the backhaul resources. 
%\footnote{Our experiments show that the proposed candidate cluster $\mathcal{C}_k$ can improve the performance of the high rate users as compared to setting equal $L_k$ for all the users.}.
The proposed candidate cluster $\mathcal{C}_k$ can be alternatively seen as one way to set 
a nonuniform $L_k$ for each user (as $L_k = |\mathcal{C}_k|$) through a common parameter $\eta_1$ as in (\ref{eq:CandidateCluster}).

\subsubsection{Biased Signal Strength Based Static Clustering}

An alternative method for controlling the traffic load of the BSs is to set a received signal strength 
bias $\zeta_l$ for each BS $l$ and to determine the user-centric clusters based on the biased signal strength. 
%based on which an even simpler 
%static clustering scheme than Algorithm~\ref{alg:GMRCF} is proposed in Algorithm~\ref{alg:HeuristicBias}. 
By setting a higher bias for the underloaded BSs, 
users around the overloaded BSs are prompted to connect with the underloaded BSs instead. 
This biasing idea originates from the SINR-bias technique for cell range expansion in the traditional 
BS-user association problem for heterogeneous networks \cite{Guvenc11}, where each user 
is only associated with a single BS. 
The proposed algorithm generalizes this idea to the case where each user can associate with a cluster of multiple BSs, 
and further combines the idea of biasing with the idea of using 
a received signal strength gap threshold, denoted as $\eta_2$, to determine the cluster sizes for different users, as done in (\ref{eq:CandidateCluster}).
The proposed biased signal strength based static clustering scheme is described in detail as Algorithm~\ref{alg:HeuristicBias} below.

\begin{algorithm}
Let $\zeta_l$ be the received signal strength bias from BS $l$, $l=1,2,\cdots,L$, 
the serving BS cluster for user $k$ is set as:
\begin{align*}
\mathcal{L}_k = \left\{l \in\mathcal{L} \vert ~ \max_{m} \left( s_{m,k} + \zeta_m \right) - \left( s_{l,k} + \zeta_l \right) \leq \eta_2 \right\}
\end{align*}
$k=1,2,\cdots,K$, where $s_{m,k}$ is the received signal strength from BS $m$ to user $k$.
\caption{Biased Signal Strength Based Static Clustering}
\label{alg:HeuristicBias}
\end{algorithm}

\section{Simulation Results}

\begin{table}[t]
\centering
\caption{Simulation Parameters.}
\label{table:system-parameter}
\begin{tabular}{|c|c|}
\hline 
Cellular  & Hexagonal  \\
     Layout        &  $7$-cell wrapped-around \\\hline
Channel bandwidth & $10$ MHz    \\ \hline
Distance between cells  &  $0.8$ km \\ \hline
% Num. of users/cell  &  $30$   \\ \hline
Num. of (macro-BSs, pico-BSs, users)$/$cell  &  $(1, 3, 30)$   \\ \hline
%Number of pico-BSs/cell  &  $3$   \\ \hline
Num. of antennas$/$(macro-BS, pico-BS, user)  &   $(4, 2, 2)$ \\ \hline
%Number of Tx antennas/pico-BS  &   $2$ \\ \hline
%Num. of Rx ant's/user  &  $2$  \\ \hline
Max. Tx power for (macro-BS, pico-BS)   &  $(43, 30)$ dBm \\ \hline
%Max. Tx Power at pico-BS   &  $30$ dBm \\ \hline
 Antenna gain & $15$ dBi \\ \hline
 Background noise  & $-169$ dBm/Hz \\ \hline
 Path loss from macro-BS to user & $128.1+ 37.6 \log_{10}(d)$ \\ \hline
Path loss from pico-BS to user & $140.7+ 36.7 \log_{10}(d)$ \\ \hline
Log-normal shadowing & $8$ dB \\ \hline
Rayleigh small scale fading & $0$ dB  \\ \hline
Reweighting function parameter & $\tau = 10^{-10}$ \\ \hline
\end{tabular}
\end{table}

\begin{figure}[t]
  \centering
  \includegraphics[width= 0.45\textwidth]{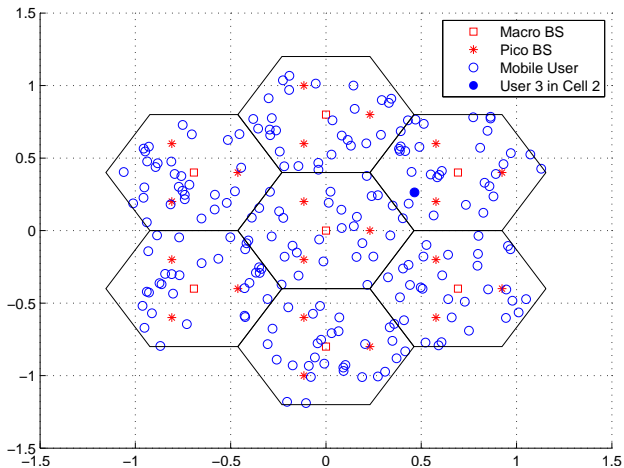}
\caption{7-cell wrapped around two-tier heterogeneous network.}
\label{fig:HetNet}
\end{figure}

In this section, numerical simulations are conducted to show the effectiveness of the proposed algorithms. 
We consider a 7-cell wrapped-around two-tier heterogeneous network with the simulation parameters listed in Table~\ref{table:system-parameter}. 
Each cell is a regular hexagon with a single macro-BS located at the center and 3 pico-BSs equally separated in space as illustrated in 
Fig.~\ref{fig:HetNet}. 
To simplify the discussion, we set all the macro-BSs to have equal backhaul constraints and likewise for the pico-BSs. 
The backhaul constraints are denoted as $(C_{\text{macro}},C_{\text{pico}})$ respectively. 
%and correspondingly denote the proposed algorithm as ``Proposed Algorithm $(C_{\text{macro}},C_{\text{pico}})$Mbps''. 
The proposed algorithms are simulated under the same power constraints listed in Table~\ref{table:system-parameter} but 
with various sets of $(C_{\text{macro}},C_{\text{pico}})$ backhaul constraints.

%The computational complexity of the proposed Algorithm~\ref{alg:LCSparse}
%mainly comes from the optimal transmit beamformer design in Step 3, which is reformulated as a SOCP and solved by the CVX. 
%The complexity to solve SOCP using interior-point method is approximately $\mathcal{O}((KLM)^3)$ \cite{Ye98}. 
%In what follows, we propose two more techniques, \emph{iterative link removal} and \emph{iterative user pool shrinking}, to further improve the efficiency of the proposed algorithm in each iteration. 
%The former aims at reducing the number of potential transmit antennas $LM$ serving each user while the latter is intended to decrease the total number of users $K$ to be considered in each iteration. 

%\subsection{Iterative Link Removal}

\subsection{Dynamic BS Clustering}

\begin{figure}[t]
  \centering
  \includegraphics[width= 0.45\textwidth]{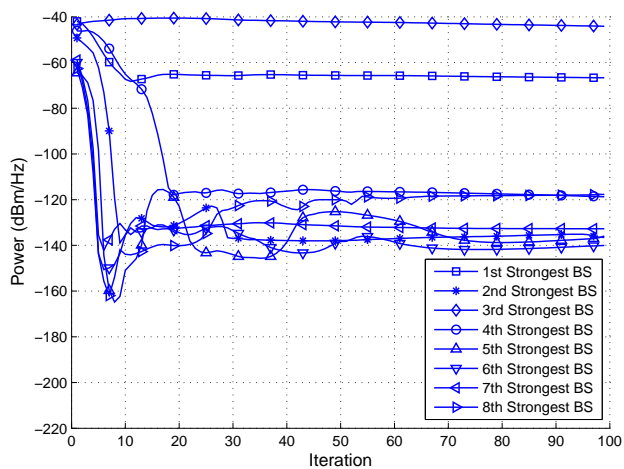}
\caption{Power evolutions of the strongest 8 BSs for user 3 in cell 2, $(C_{\text{macro}},C_{\text{pico}}) = (245, 70)$ Mbps, $\alpha_k = 1, \forall k, L_c = 8$.}
\label{fig:PowerDistUser3Cell2}
\end{figure}

\begin{figure}[t]
  \centering
  \includegraphics[width= 0.45\textwidth]{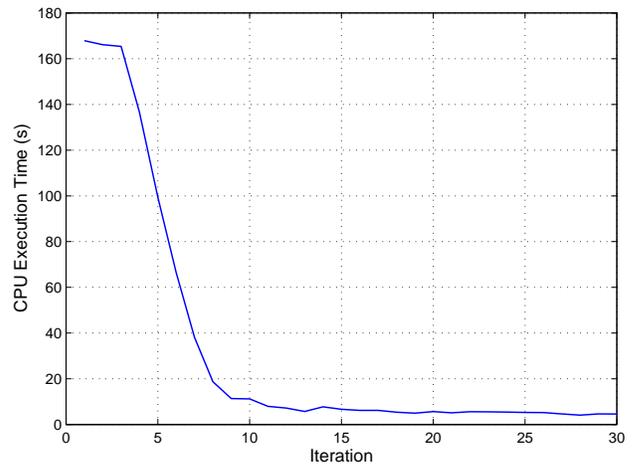}
\caption{CPU execution time needed for each iteration in a Linux x86\_64 machine with 2.3 GHz CPU and 2 GB RAM 
under $(C_{\text{macro}},C_{\text{pico}}) = (245, 70)$ Mbps, $\alpha_k = 1, \forall k, L_c = 8$.}
\label{fig:CPUTime}
\end{figure}

\begin{figure}[t]
  \centering
  \includegraphics[width= 0.45\textwidth]{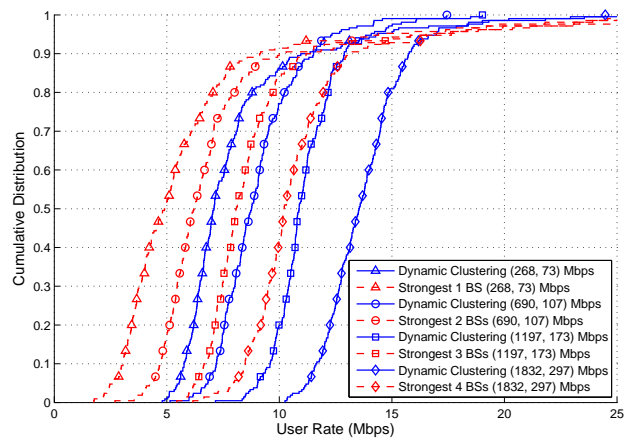}
\caption{Cumulative distribution function of user data rate comparison with $L_c = 8$ and proportionally fair scheduling.}
\label{fig:CDF}
\end{figure}

\begin{figure}[t]
  \centering
  \includegraphics[width= 0.45\textwidth]{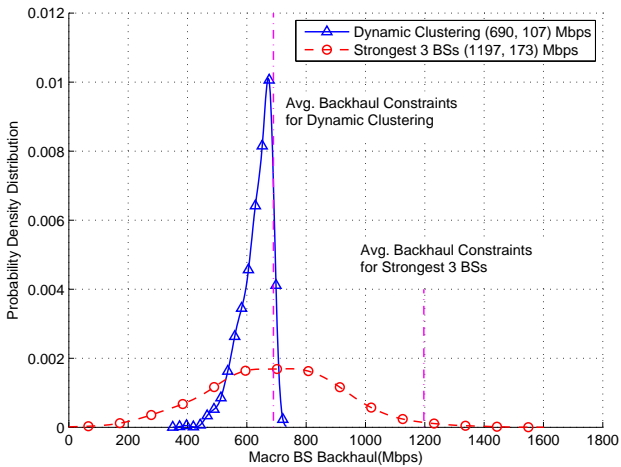}
\caption{Probability density distribution of maco-BS backhaul consumption, where the straight magenta dashed line represents
the position of the corresponding $C_{\text{macro}}$.}
\label{fig:macropdf}
\end{figure}

\begin{figure}[t]
  \centering
  \includegraphics[width= 0.45\textwidth]{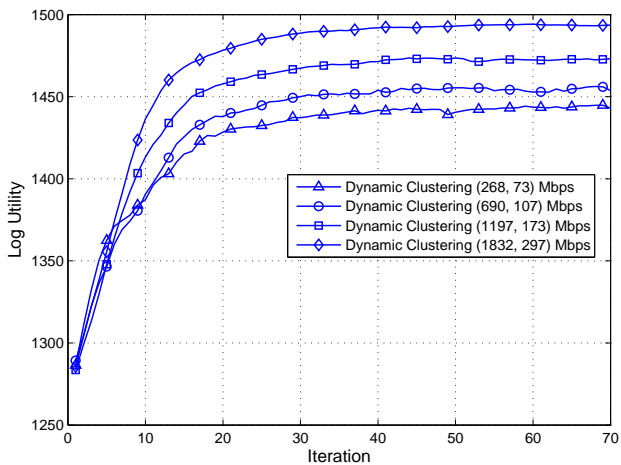}
\caption{Convergence behavior of $\sum_k \log(\bar{R}_k)$ for Algorithm~\ref{alg:LCSparse} under dynamic clustering, where $\bar{R}_k$ is the long term average rate for user $k$ and $L_c = 8$.}
\label{fig:utility}
\end{figure}

\begin{table}
\centering
\caption{50th percentile user data rate comparison.}
\begin{tabular}{ c   c c }
  \toprule
   & 50th Percentile Rate & Gain\\
  \hline
  
  Strongest 1 BS (268, 73) Mbps   & 4.9 Mbps &\multirow{2}{*}{$44.9\%$} \\ 
  Dynamic Clustering (268, 73) Mbps  & 7.1 Mbps & \\  [1.5ex]

  Strongest 2 BSs (690, 107) Mbps   & 6.2 Mbps &\multirow{2}{*}{$41.9\%$} \\ 
  Dynamic Clustering (690, 107) Mbps  & 8.8 Mbps & \\  [1.5ex]

  Strongest 3 BSs (1197, 173) Mbps  & 8.1 Mbps &\multirow{2}{*}{$34.6\%$} \\ 
  Dynamic Clustering (1197, 173) Mbps  &  10.9 Mbps &\\  [1.5ex]

   Strongest 4 BS (1832, 297) Mbps  & 10.2 Mbps &\multirow{2}{*}{$33.3\%$} \\ 
  Dynamic Clustering (1832, 297) Mbps  & 13.6 Mbps &  \\[1.5ex]
%		Fixed Clustering  & (453, 453)Mbps &  & \\
  \bottomrule  
\end{tabular}
\label{tb:RateComp}
\end{table}

We first evaluate the performance of the proposed Algorithm~\ref{alg:LCSparse} under dynamic BS clustering.
As indicated earlier, instead of considering all the $L$ BSs in the entire network as candidates for serving each user, in simulations we only consider the strongest $L_c$ ($L_c \leq L$) BSs around each user as its candidate cluster.
To illustrate how the sparse beamforming vector $\mathbf{w}_k$ for each user is formed using Algorithm~\ref{alg:LCSparse}, 
we plot in Fig.~\ref{fig:PowerDistUser3Cell2} the power evolutions of the strongest 8 BSs for the third user in the second cell as an example. As we can see, after around 20 iterations only the first and third strongest BSs maintain a reasonable transmit power level. They eventually form the cluster to serve user 3 in cell 2. 
With the proposed iterative link removal technique and by setting the threshold to be $-100$ dBm/Hz, 
Algorithm~\ref{alg:LCSparse} can narrow down the candidate BSs to
only the strongest 4 BSs after the 5th iteration, and to the (1st, 3rd, 4th) strongest BSs after the 8th iteration, and finally to 
the (1st, 3rd) strongest BSs after the 17th iteration.

To demonstrate the effectiveness of the proposed iterative link removal and iterative user pool shrinking techniques
in improving the efficiency of Algorithm~\ref{alg:LCSparse},
we plot in Fig.~\ref{fig:CPUTime} the CPU execution time needed for each iteration in Algorithm~\ref{alg:LCSparse}.
As we can see, the per-iteration execution time for Algorithm~\ref{alg:LCSparse}
drops dramatically from around 170 seconds to about 5 seconds within
20 iterations. This is due to the continually shrinking candidate cluster size and user scheduling pool. 
For instance, at the 20th iteration, the average candidate cluster size for each user is 1.96 and 
the number of remaining users in the scheduling pool is 54, which are only about $1/4$ of 
the original cluster size $L_c = 8$ and the total number of users $K=210$, respectively.

In Fig.~\ref{fig:CDF}, we compare the cumulative distributions of the 
long-term average user rates between Algorithm~\ref{alg:LCSparse} with dynamic clustering
and the baseline scheme where each user is served by the strongest
$S$ BSs ($S=1,2,3,4$). %, denoted as ``Strongest $S$ BSs''. 
We first run the baseline schemes using WMMSE algorithm without explicit backhaul constraints as in \cite{Kaviani12} and 
evaluate the corresponding backhaul requirement for each BS afterwards. 
We then set the explicit backhaul constraints $(C_{\text{macro}},C_{\text{pico}})$ in Algorithm~\ref{alg:LCSparse} 
to be the average backhaul requirements over the macro-BSs and the pico-BSs respectively computed 
from the baseline. 
The baseline is denoted as ``Strongest $S$ BSs $(C_{\text{macro}},C_{\text{pico}})$ Mbps'' while
the proposed Algorithm~\ref{alg:LCSparse} is denoted as ``Dynamic Clustering $(C_{\text{macro}},C_{\text{pico}})$ Mbps''.
Each curve in Fig.~\ref{fig:CDF} is obtained by iteratively simulating the corresponding scheme 
with fixed user priority weights $\alpha_k$'s and updating the weights as $\alpha_k = 1/\bar{R}_k$ according to the proportional fairness criterion, where $\bar{R}_k$ is the long term average data rate for user $k$.

As we can see from Fig.~\ref{fig:CDF}, by optimizing with explicit backhaul constraints, Algorithm~\ref{alg:LCSparse} achieves significant performance gain. 
We list the 50th percentile user rate comparison between Algorithm~\ref{alg:LCSparse} and the baseline in Table~\ref{tb:RateComp}.
For instance, around $35\%$ 
improvement is obtained for the 50th percentile user as
compared with the baseline where each user is served by the strongest 3 BSs.
Note that since the baseline algorithm connects each user to an equal number of neighboring BSs, it inevitably favors 
high-rate users. In contrast, the proposed Algorithm~\ref{alg:LCSparse} can select clusters for each user 
adaptively, and in particular can choose a larger cluster for 
the low-rate users, thus achieving a significant overall gain from a 
network utility perspective.

We also see from Fig.~\ref{fig:CDF} that the ``Dynamic Clustering $(690, 107)$ Mbps'' scheme
shows better performance as compared to the baseline where each user is served by the strongest 3 BSs 
as in the ``Strongest 3 BSs $(1197, 173)$ Mbps'' scheme, while only requires about $60\%$ of backhaul.
To further investigate the reason behind this significant reduction in backhaul 
consumption, we plot the probability density
distribution (pdf) of the backhaul consumption in each time slot for all the 7 macro-BSs in Fig.~\ref{fig:macropdf}.
As we can see, the macro-BS backhaul consumption for Algorithm~\ref{alg:LCSparse} 
has a peak around its constraint\footnote{The slight excess over the backhaul constraint in the pdf curve is due to the 
embedded pdf estimation function in MATLAB, which interpolates more points at the boundary to smooth the estimated curve.} 
and is highly concentrated. However, the backhaul consumption 
for the baseline is more spread out, whose peak point is only about half of its backhaul requirement. 
Similar phenomenon also occurs in the backhaul consumption for the pico-BSs.
The inefficient usage of the backhaul resource in the baseline is due to fixed clustering and the lack of explicit backhaul constraints.
In contrast, the proposed dynamic clustering with explicit backhaul constraints can adaptively form the clusters 
for each user according to the available backhaul budget, which also
has an effect of balancing the data traffic.
%Therefore, explicit backhaul constraints not only help the system control the backhaul consumption but also lead to the fully utilization of the backhaul
%resource and improvement in the system performance.

Although a rigorous theoretical proof for the convergence of the proposed Algorithm~\ref{alg:LCSparse} is not yet available, 
we plot in Fig.~\ref{fig:utility} the log utility evolutions for Algorithm~\ref{alg:LCSparse} under the same set of backhaul constraints
as in Fig.~\ref{fig:CDF}. 
As we can see, the proposed Algorithm~\ref{alg:LCSparse} with dynamic BS clustering
converges in roughly 40-50 iterations under each of the $(C_{\text{macro}},C_{\text{pico}})$ settings.

\subsection{Static BS Clustering}

\begin{figure}[t]
  \centering
  \includegraphics[width= 0.45\textwidth]{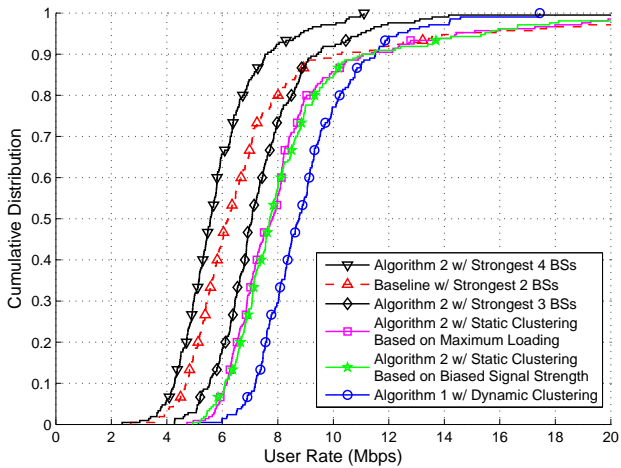}
\caption{Cumulative distribution function of user data rate comparison with backhaul constraint $(C_{\text{macro}},C_{\text{pico}}) = (690, 107)$ Mbps under proportionally fair scheduling.}
\label{fig:Strongest234}
\end{figure}

%
%\begin{figure}[t]
  %\centering
  %\includegraphics[width= 7 cm]{Static.eps}
%\caption{User rates comparison under $(C_{\text{macro}},C_{\text{pico}}) = (690, 107)$Mbps and 
%$\alpha_k$ updated according to proportional fairness criterion.}
%\label{fig:Static}
%\end{figure}

\begin{figure}[t]
  \centering
  \includegraphics[width= 0.45\textwidth]{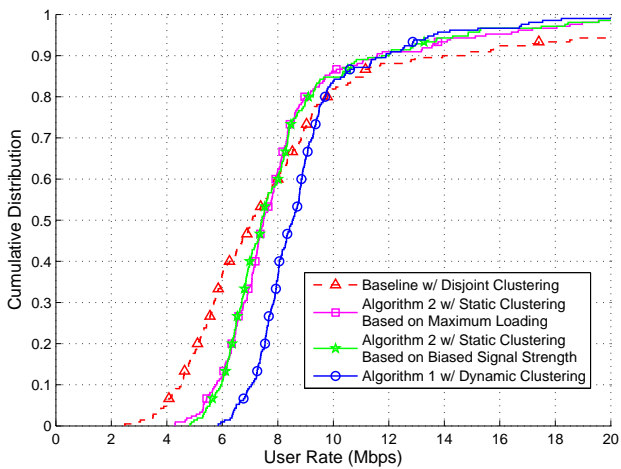}
\caption{Cumulative distribution function of user data rate comparison with backhaul constraint $(C_{\text{macro}},C_{\text{pico}}) = (453, 453)$ Mbps under proportionally fair scheduling.}
\label{fig:SingleCell}
\end{figure}

We now evaluate the performance of Algorithm~\ref{alg:LCSparseStatic} under static BS clustering schemes.
In Fig.~\ref{fig:Strongest234}, we set the ``Strongest 2 BSs'' scheme from 
Fig.~\ref{fig:CDF} as the baseline and set its corresponding backhaul requirement 
$(C_{\text{macro}},C_{\text{pico}}) = (690, 107)$ Mbps as the explicit backhaul constraint for Algorithm~\ref{alg:LCSparseStatic} 
under different static BS clustering schemes. 
We first compare the baseline with Algorithm~\ref{alg:LCSparseStatic} in which each user is associated with 
the strongest 3 and 4 BSs, respectively.
As we can see, by connecting each user to an
additional BS (strongest 3 BSs), the proposed Algorithm~\ref{alg:LCSparseStatic} with explicit backhaul constraints
improves the overall network utility by
sacrificing the high-rate users to improve the performance of the low-rate users
while still keeping the same backhaul consumption. 
However, if each user is further connected to one more additional BS (strongest 4 BSs), 
then the overall performance is even worse than the case where each user is connected to the strongest 2 BSs. This is 
because the BSs are now overloaded under the given backhaul capacity constraints and are forced to schedule only the low-rate users.
This illustrates the importance of choosing BS cluster size. 

We also plot in Fig.~\ref{fig:Strongest234} the performance of Algorithm~\ref{alg:LCSparseStatic} under the proposed static 
clustering schemes as listed in Algorithm~\ref{alg:GMRCF} and Algorithm~\ref{alg:HeuristicBias}, respectively, again with 
$(C_{\text{macro}},C_{\text{pico}}) = (690, 107)$ Mbps. 
In the maximum loading based static clustering scheme, 
the maximum number of users each macro-BS and each pico-BS can support is set to be 
$K_{\text{macro},max} = 70, K_{\text{pico},max} = 10$ and the received signal strength gap is set as $\eta_1 = 14$ dB. 
In the biased signal strength based static clustering scheme, 
the received signal strength bias is set as $\zeta_{\text{macro}} = 0$ dB for the 
macro-BSs and $\zeta_{\text{pico}} = 6$ dB for the pico-BSs and the received signal strength gap is set as $\eta_2 = 12$ dB. 
As we can see, the two proposed clustering schemes have similar performance, 
and both can achieve a significant portion of the performance gain that the dynamic clustering scheme achieves over the 
baseline of simply choosing the strongest BSs. 
For instance, compared with the baseline at the $50$th percentile user rate, 
both of the proposed static clustering schemes show around $26\%$ performance improvement, 
while the proposed Algorithm~\ref{alg:LCSparse} under dynamic clustering achieves $41.9\%$ gain as listed in Table~\ref{tb:RateComp}.

In Fig.~\ref{fig:SingleCell}, we compare the proposed static clustering schemes 
with a different baseline. The baseline scheme is chosen to be a disjoint clustering scheme where each user is jointly served by 
the 4 BSs, i.e. 1 macro-BS and 3 pico-BSs, within its cell. 
Similar to Fig.~\ref{fig:CDF}, we first run the disjoint clustering scheme using WMMSE algorithm \cite{Kaviani12}, then set
the backhaul constraints for the proposed algorithms as the average backhaul consumptions evaluated from the disjoint clustering scheme. 
Since the macro-BS and the pico-BSs in each cell are associated with the same set of users, the average backhaul consumptions for the 
macro-BSs and for the pico-BSs are equal in the disjoint clustering scheme, i.e. $(C_{\text{macro}},C_{\text{pico}}) = (453, 453)$ Mbps. 
In this case, the setting of $K_{\text{macro},max} = 50, K_{\text{pico},max} = 20$ and $\eta_1 = 20$ dB provides the best performance 
for the proposed static clustering scheme in Algorithm~\ref{alg:GMRCF}, 
while the setting of $\zeta_{\text{macro}} = 0$ dB, $\zeta_{\text{pico}} = 14$ dB and 
$\eta_2 = 12$ dB provides the best performance for the proposed static clustering scheme in Algorithm~\ref{alg:HeuristicBias}. 
As we can see from Fig.~\ref{fig:SingleCell}, since the cluster-edge users suffer from considerable inter-cluster interference, the 
disjoint clustering scheme has a substantial number of low-rate users, while the user-centric clustering schemes 
proposed in this paper can effectively improve the performance of the low-rate users.
For example, for the $10$th percentile user, both of the proposed static clustering schemes improve the rate over the disjoint clustering 
scheme by about $32\%$, while the proposed Algorithm~\ref{alg:LCSparse} under dynamic clustering achieves around $57\%$ improvement.

\section{Conclusion}

This paper proposes an $\ell_0$-norm formulation for the per-BS backhaul constraint in a downlink C-RAN system.
By taking advantage of the $\ell_1$-norm reweighting technique, the nonconvex per-BS backhaul constraint is approximated as a 
convex weighted per-BS power constraint. 
This approximation allows us to use a generalized WMMSE approach to solve the WSR maximization problem under two different cases, 
depending on whether BS clustering is dynamic or static over different time-frequency slots.
In the former case, we propose a joint user-centric clustering, user scheduling and beamforming design algorithm; in the latter case we fix the BS cluster for each user and jointly optimize the user scheduling and beamforming vectors.  
We also provide two static clustering formation algorithms, which can effectively take into account 
both the traffic load of each BS and the channel condition for each user. 
Simulation results show that with explicit per-BS backhaul constraints, 
the proposed dynamic clustering scheme is able to significantly 
improve the system performance over the naive clustering schemes, while the proposed static clustering schemes 
can already achieve a substantial portion of the performance gain.

%\balance

% conference papers do not normally have an appendix

% use section* for acknowledgement
%\section*{Acknowledgment}
%
%
%The authors would like to thank...

% trigger a \newpage just before the given reference
% number - used to balance the columns on the last page
% adjust value as needed - may need to be readjusted if
% the document is modified later
%\IEEEtriggeratref{8}
% The "triggered" command can be changed if desired:
%\IEEEtriggercmd{\enlargethispage{-5in}}

% references section

% can use a bibliography generated by BibTeX as a .bbl file
% BibTeX documentation can be easily obtained at:
% http://www.ctan.org/tex-archive/biblio/bibtex/contrib/doc/
% The IEEEtran BibTeX style support page is at:
% http://www.michaelshell.org/tex/ieeetran/bibtex/
%\bibliographystyle{IEEEtran}
% argument is your BibTeX string definitions and bibliography database(s)
%\bibliography{IEEEabrv,../bib/paper}
%
% <OR> manually copy in the resultant .bbl file
% set second argument of \begin to the number of references
% (used to reserve space for the reference number labels box)
%\begin{thebibliography}{1}
%
%\bibitem{IEEEhowto:kopka}
%H.~Kopka and P.~W. Daly, \emph{A Guide to \LaTeX}, 3rd~ed.\hskip 1em plus
  %0.5em minus 0.4em\relax Harlow, England: Addison-Wesley, 1999.
%
%\end{thebibliography}

\bibliographystyle{IEEEtran}
%\bibliography{strings,refs}

\bibliography{IEEEabrv,myref}

\end{document}